\documentclass[accepted]{uai2023} 

\usepackage[american]{babel}

\usepackage{natbib} 
\usepackage{mathtools} 
\usepackage{booktabs} 
\usepackage{tikz} 

\usepackage{algorithm}
\usepackage{algpseudocode}
\usepackage{amsmath}
\usepackage{stfloats}
\usepackage{color}
\usepackage{graphicx}
\usepackage{subfigure}
\usepackage{longtable}
\usepackage{makecell}
\usepackage{enumitem}
\usepackage{amssymb}
\usepackage{multirow}
\usepackage{svg} 
\usepackage{amsthm}
\usepackage{diagbox}
\newtheorem{theorem}{Theorem}

\newtheorem{definition}{Definition}

\title{Practical Privacy-Preserving Gaussian Process Regression via Secret Sharing}

\author[1,2]{Jinglong Luo}
\author[2]{Yehong Zhang\thanks{Corresponding author}}
\author[2]{Jiaqi Zhang}
\author[2]{Shuang Qin}
\author[2]{Hui Wang}
\author[2]{Yue Yu}
\author[1,2]{Zenglin Xu$^*$}
\affil[1]{%
    Harbin Institute of Technology, Shenzhen, China
}
\affil[2]{%
    Peng Cheng Laboratory, Shenzhen, China
}

\begin{document}
\maketitle

\begin{abstract}
  \emph{Gaussian process regression} (GPR) is a non-parametric model that has been used in many real-world applications that involve sensitive personal data (e.g., healthcare, finance, etc.) from multiple data owners.
To fully and securely exploit the value of different data sources, this paper proposes a privacy-preserving GPR method based on \emph{secret sharing} (SS), a \emph{secure multi-party computation} (SMPC) technique.
In contrast to existing studies that protect the data privacy of GPR via homomorphic encryption, differential privacy, or federated learning, our proposed method is more practical and can be used to preserve the data privacy of both the model inputs and outputs for various data-sharing scenarios (e.g., horizontally/vertically-partitioned data).
However, it is non-trivial to directly apply SS on the  conventional GPR algorithm, as it includes some operations whose accuracy and/or efficiency have not been well-enhanced in the current SMPC protocol. To address this issue, we derive a new SS-based exponentiation operation through the idea of ``confusion-correction'' and construct an SS-based matrix inversion algorithm based on Cholesky decomposition.
More importantly, we theoretically analyze the communication cost and the security of the proposed SS-based operations. 
Empirical results
show that our proposed method can achieve reasonable accuracy and efficiency under the premise of preserving data privacy.
\end{abstract}

\section{Introduction}\label{sec:intro}
\emph{Gaussian process regression} (GPR) \citep{Rasmussen2006,gp1,gp3,gp2,zhang2016near,ZheXCQP15} is a Bayesian non-parametric model that has been widely used in various real-world applications such as disease progression prediction \citep{ortmann2019automated,shashikant2021gaussian}, traffic prediction \citep{chen2015gaussian}, and finance \citep{yang2015gaussian}, etc.
In practice, the data of the above applications may belong to different parties and cannot be shared directly due to the increasing privacy concerns in the \emph{machine learning} (ML) community.
%
%
For example, two hospitals that have a small amount of patient data would like to jointly construct a high-quality GPR model for better disease progression prediction. However, such data usually contain patients' personal information and cannot be shared between hospitals due to legal regulations. In addition, when other hospitals or patients consider to use the constructed GPR model for diagnosis, privacy leakage of the personal feature (i.e., test input) and the diagnostic result (i.e., model output) is also a concern. 
Similarly, in finance, a bank that owns users' financial behaviors (e.g., income, credit, etc.) may hope to build a GPR model for risk control prediction by exploiting the users' consuming behaviors from the e-commerce companies.
Obviously, neither the financial nor the consuming behaviors of the users should be shared directly due to their high information privacy.

\begin{figure*}[t]
	\centering
	\begin{tabular}{ccc}
		\hspace{-2mm}\includegraphics[width=0.3\textwidth]{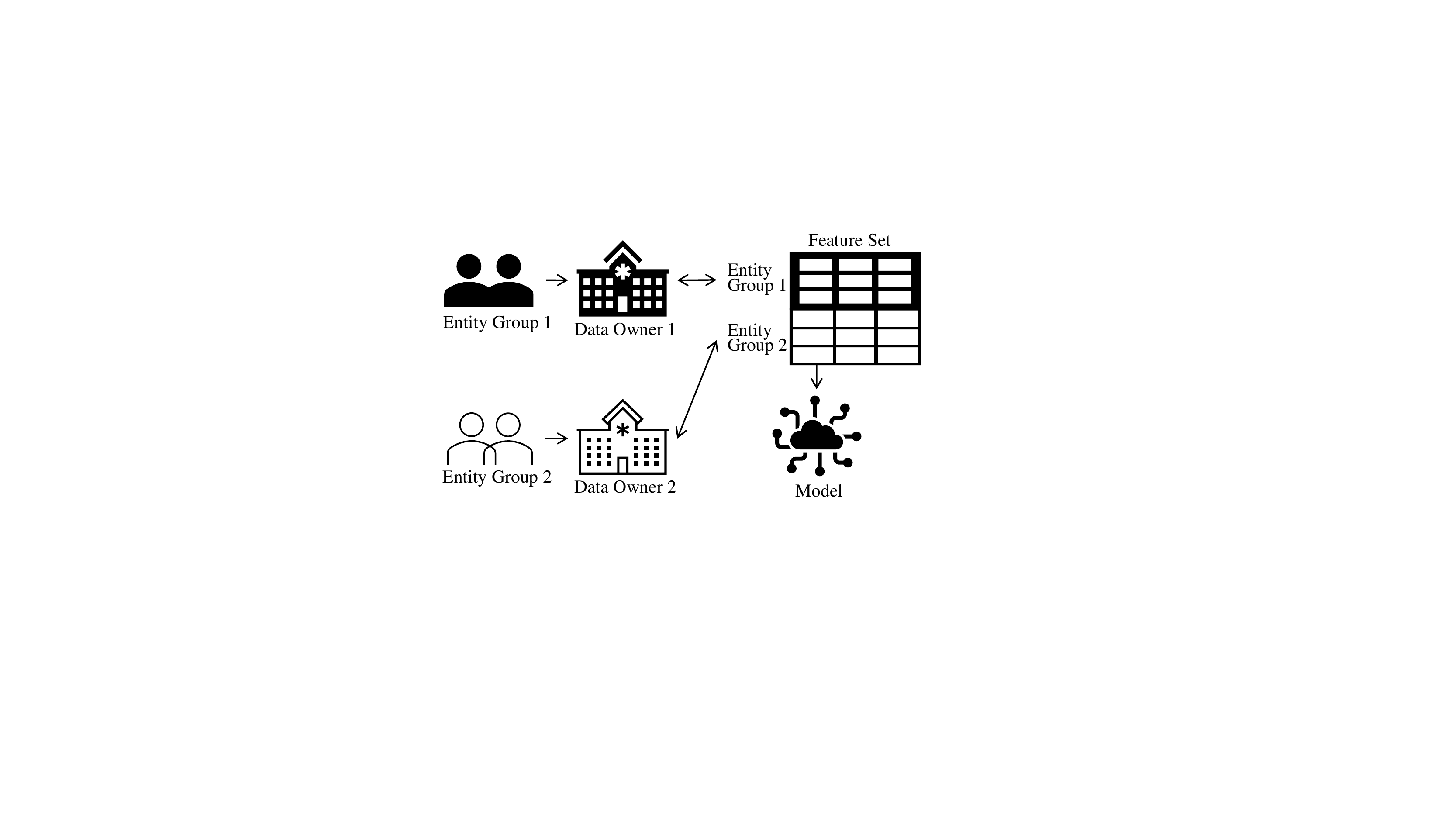} & \hspace{10mm}\includegraphics[width=0.25\textwidth]{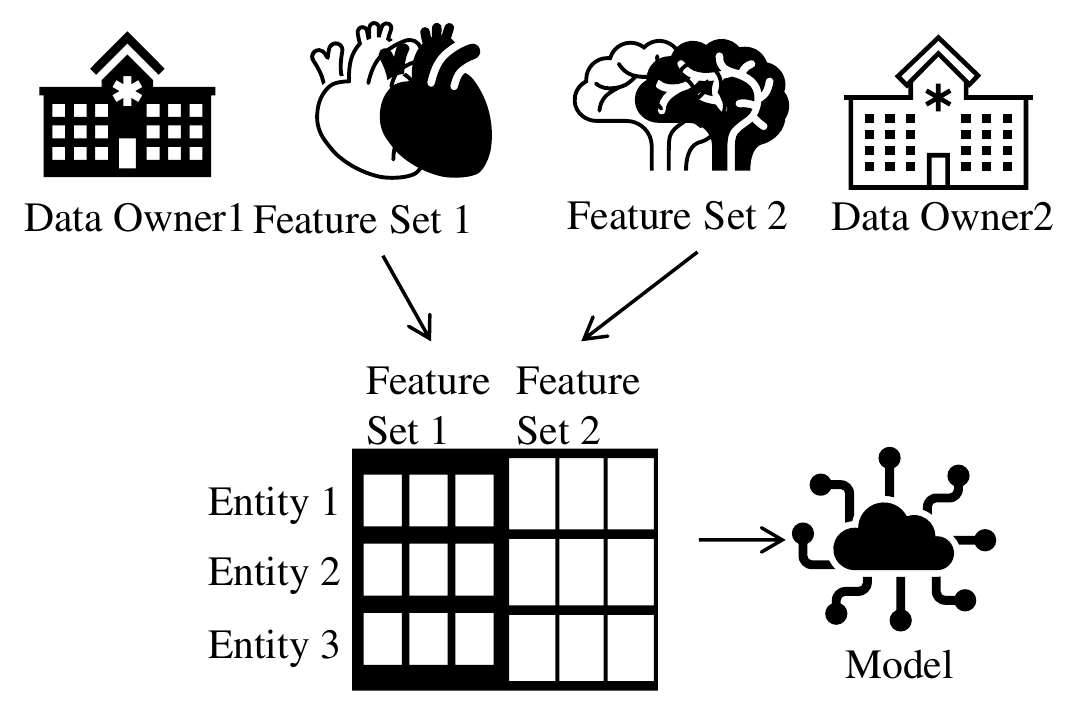} & \hspace{10mm}\includegraphics[width=0.18\textwidth]{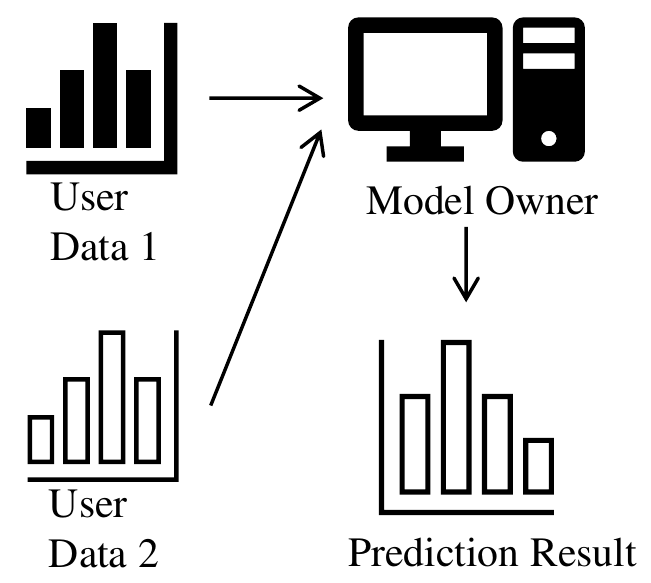}\\
		\hspace{-2mm}(a) Horizontal data-sharing & \hspace{10mm}(b) Vertical data-sharing & \hspace{7mm}(c) Prediction data-sharing\\
	\end{tabular}
	\caption{Diagrams of different data-sharing scenarios.}
	\label{fig:datashare}
\end{figure*}

The need of information sharing in the above examples has motivated the development of a practical GPR model that can preserve the data privacy of both the model inputs and outputs in three data-sharing scenarios (Fig.~\ref{fig:datashare}): (a) \emph{Horizontal data-sharing} (HDS): Each party has a set of data for different entities with the same features and share them for model construction;
(b) \emph{Vertical data-sharing} (VDS): Each party has different features of the same set of entities and shares them for model construction;
and (c) \emph{Prediction data-sharing} (PDS): A party who aims to use the constructed model needs to share his data with the model holder for prediction.
%
At present, a few privacy enhancement techniques such as \emph{fully homomorphic encryption} (FHE)~\citep{gentry2009fully},
\emph{federated learning} (FL)~\citep{konevcny2016federated},
and \emph{differential privacy} (DP)~\citep{dwork2006differential, abadi2016deep} have been exploited for avoiding privacy leakage in GPR.
However, none of them is general enough to achieve privacy-preserving GPR for all three data-sharing scenarios.
Specifically, the FHE-GPR \citep{fenner2020privacy} and FL-GPR \citep{dai2020federated,dai2021differentially,kontoudis2022fully,yue2021federated} approaches only focus on PDS and HDS scenarios, respectively. 
The DP-GPR methods \citep{kharkovskii2020private,smith2018differentially} assume all the data belong to a single party and can only protect the privacy of either the input features \citep{kharkovskii2020private} or the outputs \citep{smith2018differentially}.
See Section~\ref{sec:related} for detailed discussions.

To fully and securely exploit the value of different data sources in the aforementioned data-sharing scenarios for a GPR model,
this paper proposes to exploit the \emph{secure multi-party computation (SMPC)} \citep{Yao1986MPC} which can deal with different data-sharing scenarios with a theoretical security guarantee.
%
Among the various types of SMPC approaches \citep{evans2018mpcsurveypaper, Goldreich2019GMW,Yao1982protocolsGC}, \emph{secret sharing} (SS) \citep{shamir1979share} is exploited in this work due to its good communication efficiency and widely applications in other ML models \citep{Mohassel2017SecureML, Wagh2019SecureNN}.
As the name implies, an SS-based ML approach converts all the original operations (e.g., addition, multiplication, comparison, etc.) in the ML models (e.g., neural network) into its privacy-preserving alternatives which take secretly shared data as input and produce secretly shared results with secure information communication among parties (see Section~\ref{sec:SS} for details).

Although many SS-based operations have been developed in existing privacy-preserving ML works, they are not sufficient for constructing a privacy-preserving GPR model since the matrix inversion and exponentiation operations are essential for GPR (Section~\ref{sec:PP-GPR}) but have not been well adapted to SMPC. 
To be specific, some works \citep{knott2021crypten} designed SMPC protocols of these two operations based on approximation methods such as Newton-Raphson iteration method and Taylor expansion, which significantly reduces their accuracy and/or efficiency. 
To address this issue, this work proposes new SMPC protocols for positive-definite matrix inversion and exponentiation based on SS and integrate them into the existing SS-based operations for achieving an efficient and theoretically secure GPR model.
The specific contributions of this work include:

\begin{itemize}[leftmargin=*,itemsep=5pt,topsep=0pt,parsep=0pt,partopsep=0pt]

\item To the best of our knowledge, this is the first work that considers to protect the privacy of a GPR model via secret sharing which can be used for various data-sharing scenarios (Section~\ref{sec:PP-GPR}).


\item Based on the SS technique, we propose an efficient \emph{privacy-preserving exponentiation} algorithm through the idea of ``confusion-correction'', which is shown to be $10 \sim 70$ times faster than commonly-used approximation algorithms and can achieve theoretical correctness and security guarantees (Section~\ref{sec:ppoperators}).

\item We propose the first SS-based matrix inversion algorithm via Choseky decomposition and show that its accuracy is comparable to the plaintext algorithm with acceptable communication cost (Section~\ref{sec:ppoperators}).

\item Empirical results on two real-world datasets show that the proposed SS-based GPR algorithm can achieve accurate prediction results within a reasonable time (Section~\ref{sec:experi}).
\end{itemize}

\section{Background and notations}
\label{sec:gen_inst}

\subsection{Gaussian process regression (GPR)}\label{sec:GPR}

Let $\mathcal{X}$ denote a $d$-dimensional input domain. For each $\mathbf{x} \in \mathcal{X}$, we assume its corresponding output $y(\mathbf{x} ) \sim \mathcal{N}(f(\mathbf{x}), \sigma^2_n)$ is a noisy observation of a function $f(\mathbf{x})$ with noise variance $\sigma^2_n$. Then, the function $f(\mathbf{x})$ can be modeled using a \emph{Gaussian process} (GP), that is, every finite subset of $\{f(\mathbf{x})\}_{\mathbf{x} \in \mathcal{X}}$ follows a multivariate Gaussian distribution.
Such a GP is fully specified by its \emph{prior} mean $\mu(\mathbf{x}) \triangleq \mathbb{E}[f(\mathbf{x})]$ and covariance $k(\mathbf{x}, \mathbf{x}^\prime) \triangleq \text{cov}[f(\mathbf{x}), f(\mathbf{x}^\prime)]$ for all $\mathbf{x}, \mathbf{x} \in \mathcal{X}$.
Usually, we assume that $\mu(\mathbf{x}) = 0$ and the covariance is defined by a kernel function. One example of the widely-used kernel function is the \emph{squared exponential} (SE) kernel:
\begin{equation}\label{kernel}
k(\mathbf{x}, \mathbf{x}') \triangleq \sigma^2_s\text{exp}(-d(\mathbf{x}, \mathbf{x}^\prime)/2\ell^2)
\end{equation}
where $d(\mathbf{x}, \mathbf{x}^\prime) = ||\mathbf{x} - \mathbf{x}^\prime||^2_2$, $\ell$ is the length-scale and $\sigma_s^2$ is the signal variance.

Supposing we have a set $\mathcal{D}$ of $n$ observations: $\mathcal{D} = \{(\mathbf{x}_i, y_i)_{i=1}^n\}$ where $y_i \triangleq y(\mathbf{x}_i)$, a GPR model can 
perform probabilistic regression by providing a predictive distribution $p(f(\mathbf{x}_*)|\mathcal{D}) \triangleq \mathcal{N}(\mu_{\mathbf{x}_*|\mathcal{D}}, \sigma^2_{\mathbf{x}_*|\mathcal{D}})$ for any test input $\mathbf{x}_* \in \mathcal{X}$. Let $\mathbf{X} \triangleq (\mathbf{x}_1, \ldots, \mathbf{x}_n )^\top$ be an $n \times d$ input matrix and $\mathbf{y} = (y_1, \ldots, y_n)^\top$ be a column vector of the $n$ noisy outputs.
Then, the \emph{posterior} mean and variance of the predictive distribution $p(f(\mathbf{x}_*)|\mathcal{D})$ can be computed analytically:
%
%
\begin{equation}\label{GPpred}
\begin{array}{c}
\mu_{\mathbf{x}_*|\mathcal{D}} \triangleq \mathbf{k}^\top_* (\mathbf{K} + \sigma^2_n \mathbf{I})^{-1}\mathbf{y}\ , 
\vspace{2mm}\\
\sigma^2_{\mathbf{x}_*|\mathcal{D}} \triangleq k(\mathbf{x}_*, \mathbf{x}_*) - \mathbf{k}^\top_* (\mathbf{K} + \sigma^2_n \mathbf{I})^{-1} \mathbf{k}_* ,
\end{array}
\end{equation}
where $\mathbf{k}_* \triangleq k(\mathbf{x}_*, \mathbf{X}) =  (k(\mathbf{x}_*, \mathbf{x}_i))^n_{i = 1}$ is a column vector of $n$-dimension, $\mathbf{K} \triangleq k(\mathbf{X}, \mathbf{X}) = (k(\mathbf{x}_i, \mathbf{x}_j))_{i, j = 1, \ldots, n}$ is an $n \times n$ gram matrix, and $\mathbf{I}$ is an identity matrix of size $n$.

\subsection{Secure multi-party computation}

\emph{Secure multi-party computation} (SMPC) is a type of cryptography technique for multiple parties to jointly compute an operation $f$ without exposing the privacy of the data to any of them during the computation.
%
In this work, we adopt the \textit{semi-honest security} (also known as \emph{honest-but-curious}) model which is one of the standard security models in SMPC and has been widely used in existing privacy-preserving machine learning algorithms \citep{liu2017oblivious,Mohassel2018ABY3,Mohassel2017SecureML,ryffel2020ariann,Wagh2019SecureNN}.
In a \emph{semi-honest} security model, the parties are assumed to follow the SMPC protocol but can try to use the obtained data-sharing and intermediate results to infer the information that is not exposed to them during the execution of the protocol.
Next, we will first present a \emph{secret sharing} technique designed based on the \emph{semi-honest security} model to construct SMPC protocols and then, introduce the algebraic structure used for designing the SMPC protocal in this work.

\subsubsection{Secret Sharing}\label{sec:SS}

\emph{Secret sharing} (SS) is a technique independently proposed by \citet{shamir1979share} and \citet{blakley1979safeguarding} with its full name called $(t, m)$-threshold secret sharing schemes, where $m$ is the number of parties and $t$ is a threshold value.
The security of SS requires that any less than $t$ parties cannot obtain any secret information jointly.
As a special case of secret sharing, $(2,2)$-\emph{additive} secret sharing contains two algorithms: $Shr(\cdot)$ and $Rec(\cdot, \cdot)$.
Let $\mathcal{Z}_L$ denote the ring of integers modulo $L$ and $[\![u]\!]=([u]_0, [u]_1)$ be the additive share of any integer $u$ on $\mathcal{Z}_L$.
$Shr(u) \rightarrow ([u]_0, [u]_1)$ is used to generate the share by randomly selecting a number $r$ from $\mathcal{Z}_L$, letting $[u]_0=r$, and computing $[u]_1=(u - r)\mod L$.
Note that due to the randomness of $r$, neither a single $[u]_0$ nor $[u]_1$ can be used to infer the original value of $u$. The algorithm $Rec([u]_0, [u]_1)\rightarrow u$ is used to reconstruct the original value from the additive shares, which can be done by simply calculating $([u]_0 + [u]_1) \mod L$.  

The additive secret sharing technique has been widely used to construct SMPC protocols for ML operations (e.g., addition, multiplication, etc.) such that both the inputs and outputs of the protocol can be \emph{additive} shares of the original inputs and outputs:
$
\pi_f([inputs]_0, [inputs]_1) \rightarrow [f]_0, [f]_1
$
where $\pi_f$ denotes an SMPC protocol of the operation $f$.
To further elaborate the SS technique, we briefly introduce the SS-based multiplication protocol below, which is essential in many privacy-preserving ML algorithms and will also be widely used in this work.

\textbf{SS-based multiplication $u\cdot v$\ .} Let ${P_j}$ with $j \in \{0,1\}$ be two parties that are used to execute the SMPC protocol. Each party $P_j$ will be given one additive share $([u]_j, [v]_j)\in \mathcal{Z}_L$ of the operation inputs for $j \in \{0,1\}$. Then, the additive shares of $u\cdot v$ can be computed with Beaver-triples \citep{Beaver1991triples}: $(a,b,c)$ where $a, b \in \mathcal{Z}_{L}$ are randomly sampled from $\mathcal{Z}_{L}$ and $c = a \cdot b \mod L$. Specifically, for each $j \in \{0, 1\}$,  $P_j$ first calculates $[d]_j = [u]_j-[a]_j$ and $[e]_j = [v]_j - [b]_j$. Then, they send the $[d]_j$ and $[e]_j$ to each other and reconstruct $d= Rec([d]_0, [d]_1)$ and $e =  Rec([e]_0, [e]_1)$. Finally, the additive share of $u \cdot v$ can be computed using $[u\cdot v]_j = -jd \cdot e +[u]_j \cdot e + d \cdot [v]_j + [c]_j$.
To complete the SS-based multiplication, both parties need to spend $1$ round of two-way communication and transmit two ring elements.

The SS-based multiplication protocol is extended to matrix multiplication in the work of \citet{Mohassel2017SecureML}.
Let $\mathcal{F}_{matMul}$ denote the SS-based matrix multiplication functionality, $\mathbf{U}$ and $\mathbf{V}$ be two matrices of size $m\times n$ and $n\times k$, respectively.
The SS-based matrix multiplication $\mathcal{F}_{matMul}(\mathbf{U}, \mathbf{V})$ still requires only $1$ rounds of bidirectional communication between parties $P_0$ and $P_1$ but the transmitted ring elements are of size $(m+n)\times k$.

Unfortunately, there exist many operations (e.g., exponentiation, matrix inversion, etc.) that cannot be constructed using purely additive secret sharing on $\mathcal{Z}_L$. Some approximation methods such as Newton-Raphson iteration method and Taylor expansion have been exploited for designing additive SS-based protocols of these operations. Details of the approximation methods and other SMPC protocols can be found in the work of \citet{knott2021crypten}. 



\subsubsection{Fixed-Point Representation}

As has been shown above, the SS-based SMPC protocols are constructed in a ring of integers due to security reasons.
In practice, the ML algorithms such as GPR are usually implemented using floating-point numbers.
However, it has been shown that the SMPC protocols designed based on floating-point numbers are inefficient ~\citep{aliasgari2012secure} and fixed-point representation is a better choice. 
%
%

Specifically, the fixed-point encoding method represents all data as $l$ bits. Let $\mathcal{Q}_{<\mathcal{Z}_L,l_f>}$ be a set of fixed-point numbers with a precision of $l_f$ (i.e., $l_f$ fractional bits) mapped from $\mathcal{Z}_L$ and $L = 2^l$. 
For floating-point numbers in the range\footnote{We assume that all the numbers appeared in an ML algorithm are in this range. Appropriate $l$ and $l_f$ need to be selected for avoiding underflow and overflow issues.} $[-2^{l-l_f-1}, 2^{l-l_f-1})$, this work will first round them to the nearest fixed-point numbers in $\mathcal{Q}_{<\mathcal{Z}_L,l_f>}$ and then, map them to the integers in $\mathcal{Z}_L$ by multiplying the converted fixed-point numbers with $2^{l_f}$.
For example, given $l = 5$ and $l_f = 3$, a floating-point number $1.125123$ in $[-2,2]$ is firstly rounded to the fixed-point number $1.125$ in $\mathcal{Q}_{<\mathcal{Z}_{2^5},3>}$ and then converted to an element in $\mathcal{Z}_{2^5}$ by $(1.125 \times 2^3) \mod 2^5= 9$. Conversely, an integer $11 \in \mathcal{Z}_{2^5}$ can be converted in  $\mathcal{Q}_{<\mathcal{Z}_{2^5},3>}$ as $11/2^3 = 1.375$.
%

All the algorithms in this paper are performed on $\mathcal{Z}_L$ and $Q_{<\mathcal{Z}_L,l_f>}$.
By choosing appropriate $l$ and $l_f$, the fixed-point-based SMPC protocols can achieve a desirable compromise between efficiency and accuracy.
To ease notations, we will use lowercase letters to represent either floating-point or integer numbers and $\check{x}$ to represent a fixed-point number in $\mathcal{Q}_{<\mathcal{Z}_{L},l_f>}$ converted from $x$. The algorithm $Shr(x)$ in Section~\ref{sec:SS} will first convert $x$ to its corresponding representation in $\mathcal{Z}_L$ if the input $x$ is a floating-point number.

\section{Privacy-Preserving GPR}\label{sec:PP-GPR}

\begin{figure}
	\centering
	\includegraphics[scale=0.43]{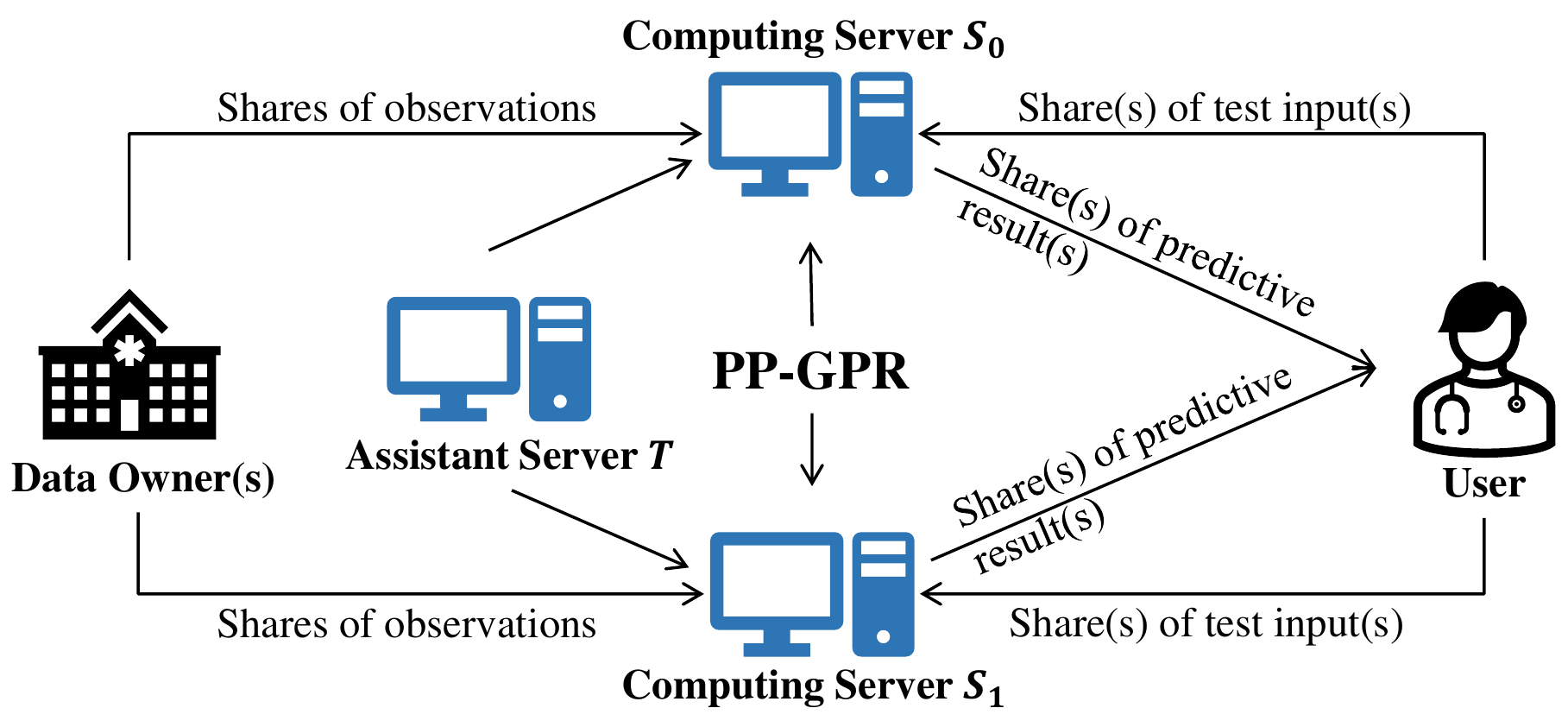}
	\caption{The overall framework of PP-GPR.}
	\label{fig:gpr}
\end{figure}

In this section, we propose to exploit the SMPC technique for constructing a \emph{privacy-preserving GPR} (PP-GPR) algorithm.
%
%
The overall framework of the \emph{Privacy-preserving GPR} (PP-GPR) model is shown in Fig.~\ref{fig:gpr}.
As can be seen, the PP-GPR adopts a three-party SMPC architecture with two computing servers and one assistant server. Let $S_0$ and $S_1$ represent the two computing servers and $T$ be the assistant server.
Each computing server takes one additive share of the data as input, performs the calculations according to the steps of the PP-GPR algorithm, and then outputs the additive share of the GPR predictive results. The assistant server is responsible for generating random numbers required during the execution of the SS-based protocols in the PP-GPR algorithm. 
%
%
%
The proposed algorithm exploits the SS-based operations for achieving privacy-preserving GPR on all three data-sharing scenarios shown in Fig.~\ref{fig:datashare}. The complete steps are illustrated in Algorithm~\ref{alg-PPGPR}.

\begin{algorithm}[t] \footnotesize
	\caption{Privacy-preserving GPR}
	\label{alg-PPGPR}
	\textbf{Setup:} The servers determine $\mathcal{Z}_L$ and $\mathcal{Q}_{<\mathcal{Z}_L,l_f>}$. The data owners convert 
	their private observations $\mathcal{D} = (\mathbf{X}, \mathbf{y})$ and predicted samples $\mathbf{x}_*$ into $([\! [\mathbf{x}_*]\!], [\![\mathbf{X}]\!], [\![\mathbf{y}]\!])$. \\
	%
	\textbf{Input:} For $j \in \{0, 1\}$, $S_j$ holds the shares $([\mathbf{x}_*]_j, [\mathbf{X}]_j, [\mathbf{y}]_j)$, and the hyperparameters $(\ell, \sigma_s^2, \sigma_n^2)$.
    \begin{algorithmic}[1]
    \State // \textbf{Model construction stage.}
    \State \qquad $[\![d(\mathbf{X},\mathbf{X})]\!] \leftarrow \mathcal{F}_{dist}([\![\mathbf{X}]\!],[\![\mathbf{X}]\!])$
  \State \qquad $[\![\mathbf{K}]\!] \leftarrow \sigma_s^2\cdot \mathcal{F}_{PPExp}([\![-d{(\mathbf{X},\mathbf{X})}/2\ell^2]\!])$
    \State \qquad $[\![Inv]\!] \leftarrow \mathcal{F}_{MatInv}([\![(\mathbf{K} + \sigma^2_n \mathbf{I})]\!])$
    \State // \textbf{Prediction stage.}
   \State \qquad $[\![d{(\mathbf{x}_*, \mathbf{X})}]\!] \leftarrow \mathcal{F}_{dist}([\![\mathbf{x}_*]\!],[\![\mathbf{X}]\!])$
   \State \qquad $[\![d{(\mathbf{x}_*,\mathbf{x}_*)}]\!] \leftarrow \mathcal{F}_{dist}([\![\mathbf{x}_*]\!],[\![\mathbf{x}_*]\!])$
   
  \State \quad // \ Compute the kernel matrices.
   \State \qquad $[\![\mathbf{k}_*]\!] \leftarrow \sigma_s^2\cdot \mathcal{F}_{PPExp}([\![-d{(\mathbf{x}_*,\mathbf{X})}/2\ell^2]\!])$
   \State \qquad $[\![k(\mathbf{x}_*,\mathbf{x}_*)]\!] \leftarrow \sigma_s^2\cdot \mathcal{F}_{PPExp}([\![-d{(\mathbf{x}_*,\mathbf{x}_*)}/2\ell^2]\!])$
    \State \quad // \ Compute the predictive mean and variance. 
    \State \qquad $[\![\mu^2_{\mathbf{x}_*|\mathcal{D}}]\!] \leftarrow \mathcal{F}_{matMul}(\mathcal{F}_{matMul}([\![\mathbf{k}^\top_*]\!],[\![Inv]\!]), [\![\mathbf{y}]\!])$
    \State \qquad $[\![\Lambda]\!] \leftarrow \mathcal{F}_{matMul}(\mathcal{F}_{matMul}([\![\mathbf{k}^\top_*]\!],[\![Inv]\!]), [\![\mathbf{k}_*]\!])$
   \State \qquad $[\![\sigma^2_{\mathbf{x}_*|\mathcal{D}}]\!] \leftarrow [\![k(\mathbf{x}_*,\mathbf{x}_*)]\!] - [\![\Lambda]\!]$
    %
    \end{algorithmic}
    \textbf{Output:} $S_j$ outputs the share $[\mu_{\mathbf{x}_*|\mathcal{D}}]_j, [\sigma^2_{\mathbf{x}_*|\mathcal{D}}]_j$ for $j \in \{0, 1\}$.
\end{algorithm}

\subsection{The Algorithm Setups}

To ensure a coherent execution of the algorithm, consensus must be reached among the servers ($S_0$, $S_1$, and $T$), data owners, and users regarding the algebraic structure to be employed. Specifically, an appropriate choice of $l$ and $l_f$ needs to be made for $\mathcal{Z}_{2^l}$ and $\mathcal{Q}{<\mathcal{Z}_{2^l},l_f>}$, respectively. Once consensus is established, the data owners and users proceed with the conversion of their private observations $\mathcal{D} = (\mathbf{X}, \mathbf{y})$ and test inputs $\mathbf{x}_*$ into shared representations denoted as $([\! [\mathbf{x}_*]\!], [\![\mathbf{X}]\!], [\![\mathbf{y}]\!])$. This conversion is accomplished using the function $Shr(\cdot)$ and each resulting share $( [\mathbf{x}_*]_j, [\mathbf{X}]_j, [\mathbf{y}]_j)$ is transmitted to the respective computing server $S_j$ for $j \in \{0, 1\}$. Afther that, the servers perform GPR's privacy-preserving model construction and inference. 

Let us consider an illustrative example to showcase the process. Supposing $l = 5$ and $l_f = 3$, the model user aims to privately predict the output of a test input $\mathbf{x}_* = (0.625, 0.375, 0.375)$. To achieve this, the user firstly converts $\mathbf{x}_*$ into $\mathcal{Z}_{2^3}$ by $\mathbf{x}_* \cdot 2^3 = (5, 3, 3)$
, independently generates random values $[\mathbf{x_*}]_0 = (6, 9, 6)$, and then calculates $[\mathbf{x}_*]_1 = \left((\mathbf{x}_* - [\mathbf{x}_*]_0) \mod 32 \right) = (31, 26, 29)$.
The computed value $[\mathbf{x}_*]_j$ is then transmitted to the computing server $S_j$ for $j \in \{0, 1\}$. In a similar manner, the data owners employ the $Shr(\cdot)$ mechanism to send all the values pertaining to their private observations $\mathcal{D}$ to the respective computing servers. 

Note that since $Shr(\cdot)$ is applied independently to each variable in $\mathbf{X}$, $\mathbf{y}$ and $\mathbf{x}_*$, the shares of the data can be computed easily no matter how the variables in $\mathbf{X}$, $\mathbf{y}$ and $\mathbf{x}_*$ are partitioned among the data owners. Therefore, the SS-based GPR algorithm can handle HDS, VDS, and PDS scenarios straightforwardly, which makes it practical enough to be used in various real-world applications.

The GPR hyperparameters $(\ell, \sigma_s^2, \sigma_n^2)$ are assumed to be known a priori and publicly shared between the computing servers. The privacy-preserving optimization of the hyperparameters will be considered in future work.

\subsection{The Algorithm Execution Steps}

Once the computing servers receive the shares of all the data and hyperparameters, they start to execute the SS-based protocols for PP-GPR and output the shares of the predictive results.
Similar to the conventional GPR, the PP-GPR algorithm contains two stages: model construction and prediction.
At the model construction stage (Lines 1-4), the servers first compute secret shares of the distance matrix $d(\mathbf{X}, \mathbf{X}) \triangleq (d({\mathbf{x}_i, \mathbf{x}_j}))_{i, j = 1, \ldots, n}$. Let $\mathcal{F}_{dist}(\mathbf{X}, \mathbf{X}')$ be the SS-based protocol for computing the shares of the distance matrix between $\mathbf{X}$ and $\mathbf{X}'$. $[\![d(\mathbf{X}, \mathbf{X})]\!] = \mathcal{F}_{dist}(\mathbf{X}, \mathbf{X}) = ([\![d({\mathbf{x}_i, \mathbf{x}_j})]\!])_{i, j = 1, \ldots, n}$ where $\mathcal{F}_{dist}$ can be constructed using conventional SS-based addition and (matrix) multiplication operations. Then, the servers compute $[\![\mathbf{K}]\!]$ by calling a \emph{privacy-preserving exponentiation} algorithm denoted as $\mathcal{F}_{PPExp}$ and compute $[\![Inv]\!]=[\![(\mathbf{K} + \sigma^2_n \mathbf{I})^{-1}]\!]$ by calling a \textit{privacy-preserving matrix-inverse} algorithm $\mathcal{F}_{MatInv}$ with the inputs $[\![\mathbf{K}]\!]$ and $\sigma^2_n$. The design of $\mathcal{F}_{PPExp}$ and $\mathcal{F}_{MatInv}$ will be discussed later in Section~\ref{sec:ppoperators}. 

At the prediction stage (Lines 5-14), the servers first compute $[\![\mathbf{k}_* ]\!]$ and $[\![k({\mathbf{x}_*, \mathbf{x}_*})]\!]$ by calling $\mathcal{F}_{dist}$ and $\mathcal{F}_{PPExp}$ and then, obtain the shares of the predictive mean $[\![\mu_{\mathbf{x}_*|\mathcal{D}}]\!] 
= [\![\mathbf{k}^\top_*]\!][\![Inv]\!][\![\mathbf{y}]\!]$ and variance $[\![\sigma^2_{\mathbf{x}_*|\mathcal{D}}]\!] = [\![\mathbf{k}^\top_*]\!]-[\![k_{\mathbf{x}_*, \mathbf{X}}]\!][\![Inv]\!][\![\mathbf{k}_*]\!]$ according to \eqref{GPpred}
by calling $\mathcal{F}_{matMul}$\ .

\section{Privacy-preserving operation construction}
\label{sec:ppoperators}

As has been shown in Section~\ref{sec:PP-GPR}, the privacy-preserving exponentiation $\mathcal{F}_{PPExp}$ and matrix inversion $\mathcal{F}_{MatInv}$ are essential for the PP-GPR algorithm. In this section, we will analyze the issues of existing methods for constructing these two operations, introduce the proposed algorithms, and analyze their computational complexity.  




\subsection{Privacy-preserving exponentiation}\label{sec:ppexp}
As aforementioned in Section~\ref{sec:SS}, the exponentiation cannot be constructed directly via additive SS. A commonly-used method to resolve this issue is to approximate the exponentiation using its Taylor expansion $e^u = \sum_{k = 0}^{\infty} \frac{1}{k!} u^k$ such that the exponentiation can be converted into addition and multiplication operations. However, the fact that the exponential grows much faster than the polynomial may lead to large errors in the Taylor series approximation. Although increasing the degree of the polynomial can increase the approximation accuracy, the communication cost will also increase due to the information exchange needed for SS-based multiplication.
The work of \citet{knott2021crypten} mitigated this problem via the limit approximation $e^u = \lim_{k \rightarrow \infty} (1+\frac{u}{2^k})^{2^k}$ and exploited the repeated squaring algorithm to iteratively generate polynomials of higher order quickly. However, achieving accurate approximation results with this approach still incurs high communication and computational costs.

In this work, we propose to construct a \emph{privacy-preserving exponentiation} (PP-Exp) operation by adopting the idea of \textit{confusion-correction}.
In PP-Exp, given a private number $u \in [u_{min}, 0]$, each computing server $S_j$ for $j \in \{0, 1\}$ takes the additive share $[u]_j$ of $u$ as input and then, deduces the additive shares of $e^u$ privately with some random numbers generated by $T$. 
The algorithm includes the following steps: (1) The computing servers mask the share of $u$ with a random value $r$ to obtain $[\![u-r]\!]$; (2) The computing servers jointly reveal the obfuscated value $u-r$; (3) Each computing server uses the obfuscated value to calculate the obfuscated target $e^{u-r}$; and (4) Each computing server corrects the share of $e^{u-r}$ by removing the mask and obtains the share of $e^u$. 
The pseudo-code of the PP-Exp is in Algorithm~\ref{alg-exp}.

Note that Algorithm~\ref{alg-exp} considers only negative input $u$ since the commonly used kernel function (e.g., \eqref{kernel}) of GPR involves only exponentiation of negative values. According to the range of $u$, the proposed PP-Exp can achieve correctness and security by selecting appropriate $[-\check{r}_{max}, \check{r}_{max})$ and $l_f$ as will be discussed later.

\begin{algorithm}[t]\footnotesize
	\caption{Privacy-preserving exponentiation ($\mathcal{F}_{PPExp}$)}
	\label{alg-exp}
    \textbf{Setup.} The servers determine $\mathcal{Z}_L$,  $\mathcal{Q}_{<\mathcal{Z}_L,l_f>}$, the range $[u_{min}, 0]$ of input $u$, and the range $[-\check{r}_{max}, \check{r}_{max})$.\\
    \textbf{Input.} $S_0$ holds the share $[u]_0$; $S_1$ holds the share $[u]_1$.
	\begin{algorithmic}[1]
        \State // \textbf{Offline phase executed on assistant server $T$:}
        \State \quad Draw $\check{r}$ in the range $[-\check{r}_{max}, \check{r}_{max})$ randomly
        \State \quad $r \leftarrow \check{r} \cdot 2^{l_f}$
        \Comment $\check{r} \in \mathcal{Q}_{<\mathcal{Z}_L,l_f>}$
        \State \quad  Generate $([r]_0$, $[r]_1) \in \mathcal{Z}_L$
		\State \quad Calculate $e^{-\check{r}}$ in $\mathcal{Q}_{<\mathcal{Z}_L,l_f>}$
		%
		\State \quad Generate $([e^{-\check{r}}]_0$, $[e^{-\check{r}}]_1) \in \mathcal{Z}_L$
		\State \quad  Send $[r]_j$ and $[e^{-\check{r}}]_j $ to $S_j$ for $j \in \{0, 1\}$
		\State // \textbf{Online phase:}
		\State \quad $S_j$ calculates $[d]_j \leftarrow [u]_j + [r]_j$ for $j \in \{0, 1\}$
		\State \quad $S_0$ and $S_1$ sends $[d]_0$ and $[d]_1$ to each other
		\State \quad $d \leftarrow Rec([d]_0 + [d]_1)$ \label{line:rec}
		\Comment Executed by both $S_0$ and $S_1$
		\State \quad $\check{d} \leftarrow d/2^{l_f}$
		\Comment Executed by both $S_0$ and $S_1$
		\State \quad Calculate $e^{\check{d}}$ in $\mathcal{Q}_{<\mathcal{Z}_L,l_f>}$
		\State \quad $S_j$ calculates $[e^u]_j \leftarrow e^{\check{d}} \cdot 2^{l_f} \cdot [e^{-\check{r}}]_j$ for $j \in \{0, 1\}$
	\end{algorithmic}
	\textbf{Output.} $S_j$ outputs the share $[e^u]_j$ for $j \in \{0, 1\}$.
\end{algorithm}

Firstly, to guarantee the correctness of the proposed PP-Exp algorithm, we need to make sure that all the fixed-point calculations (i.e., Line 5 and Lines 13-14) can not overflow or underflow. Consequently, the selected $[-\check{r}_{max}, \check{r}_{max})$ and $l_f$ need to satisfy the following relationship.
%
%
\begin{theorem}[\textbf{Correctness}] 
\label{theorem-exp}
For any number $u$ in the range $[u_{min},0]$, if $  (\check{r}_{max} - u_{min}) \log_2^e \leq l_f < \frac{l-1}{2}$,  the PP-Exp algorithm can correctly derive $([e^u]_0, [e^u]_1)$ from $([u]_0, [u]_1)$, satisfying $[e^u]_0+[e^u]_1= e^u$. 
\end{theorem}
For example, on the ring of integers $\mathcal{Z}_{2^{64}}$, assuming that the input $u$ takes values in the range $[-4,0]$ and $\check{r}$ takes values in the range $[-16, 16)$. By setting $l_f=29$, the correctness of the PP-Exp algorithm can be ensured.
See Appendix A for the proof.  


Next we will analyze the security of the PP-Exp algorithm. In Algorithm~\ref{alg-exp}, Line~\ref{line:rec} is the only step that will reconstruct $d$ in the fixed-point domain and has the risk of leaking the information of $u$.
Specifically, given the maximal range of $u$ and $\check{r}$ (i.e., $[u_{min}, 0]$ and $[-\check{r}_{max}, \check{r}_{max})$), the value of $d$ may be exploited to reduce the feasible range of $u$, which is an information leakage.
For example, supposing $u \in \{-2, -1, 0\}$, $r \in \{-1, 0, 1\}$, and $d = u+r$, one can infer that $u$ must be $-2$ or $-1$ if $d = -2$.

To formally analyze the amount of privacy leaked, we define the \emph{degree of information leakage} as follows:
\begin{definition}
Supposing $u$ is known to be an element of a finite set $\mathcal{U}$, the \emph{degree of information leakage} on $u$ is $\frac{1}{|\mathcal{U}|}$. 
\end{definition}
We consider an algorithm to be \emph{secure} if the \emph{degree of information leakage} of the input remains constant during the algorithm. Given a fixed precision $l_f$, let $m_u$ and $m_r$ be the amount of fixed-point numbers that can be represented in $[u_{min}, 0]$ and $[-\check{r}_{max}, \check{r}_{max})$, respectively.
The security of the PP-Exp algorithm satisfies the following theorem. 
\begin{theorem}[\textbf{Security}]
\label{theorem-exp-sec}
For any fixed number $u$ in the range $[u_{min}, 0]$,
the PP-Exp algorithm is secure with the probability $\frac{m_r-m_u+1}{m_r}$. The expected degree of information leakage on $u$ is $\frac{m_u + m_r -1}{m_u\cdot m_r}$.
\end{theorem}
See Appendix A for the proof.
%
Theorem~\ref{theorem-exp-sec} shows that the more the number of values of $r$ is greater than the number of values of $u$ (i.e., $m_r - m_u$ is larger), the PP-Exp has a larger probability to be secure.
Selecting a larger range of $\check{r}$ (i.e., larger $\check{r}_{max})$ can significantly reduce the degree of information leakage. However, a larger $\check{r}_{max}$ may result in larger $l_f$ and larger $l$ due to the results in Theorem~\ref{theorem-exp} and consequently, increase the amount of communication cost as will be shown in Section~\ref{sec:comp}. 


Note that the PP-Exp algorithm leaks the exact value of $u$ with probability only $\frac{2}{m_um_r}$, i.e., both $u$ and $r$ take the maximum or minimum value of the range in which they are located.
%
For example, when $l_f = 29$, supposing the input $u$ takes values in the range $[-4,0]$ and $r$ takes values in the range $[-16,16)$, we have $m_u = 2^{31}+1$ and $m_r = 2^{34}$.
The PP-Exp is secure with $\frac{7}{8}$ probability. The degree of information leakage is  $\frac{9}{2^{34}+8}$, an increase of $\frac{1}{2^{34}+8}$ over the secure one. The probability of revealing a particular value of $u$ is less than $\frac{1}{2^{64}}$. 


\subsection{Privacy-preserving matrix inversion}\label{sec:ppmi}

In existing work \citep{knott2021crypten}, the matrix inversion is approximated via \emph{Newton-Raphson} iteration which is a local optimization method such that its performance highly depends on the initial value of the algorithm. 
However, in the state of SS, we cannot know any information about the original input matrix such that it is difficult to find the initial inverted matrix that satisfies the convergence condition.

Note that $\mathbf{K}+\sigma_n^2\mathbf{I}$ is a positive definite matrix whose inversion can be computed via Cholesky decomposition $\mathbf{K}+\sigma_n^2\mathbf{I} = \mathbf{L D L}^\top$ where $\mathbf{L}$ is a lower triangular matrix and $\mathbf{D}$ is a diagonal matrix. To theoretically guarantee the security of a PP-MI algorithm, we choose to go into the Cholesky decomposition algorithm and convert all the operations to their corresponding SS-based version. Since the entire process of computing $\mathbf{L}$, $\mathbf{D}$, and their inverse involves only addition, multiplication, and division between matrix elements, we can exploit the existing SS-based addition, multiplication, division, and their composability \citep{canetti2001universally} for constructing a PP-MI algorithm.      
%
In the PP-MI, each computing server $S_j$ for $j \in \{0, 1\}$ takes one additive share $[\mathbf{U}]_j$ of matrix $\mathbf{U} \in Z^{n \times n}_{L}$ as input and deduces the share of $\mathbf{U}^{-1}$ privately.
The detailed steps and pseudo-code of the PP-MI are shown in Appendix B. 
%

To the best of our knowledge, this is the first work that implements the SS-based PP-MI via Cholesky decomposition.
A rigorous analysis of its communication cost is detailed in Section~\ref{sec:comp}. Additionally, the performance of the proposed approach is empirically demonstrated and evaluated in Section~\ref{sec:experi-ops}. 




\subsection{Communication complexity}\label{sec:comp}
In this section, we analyze the theoretical number of communication rounds and communication volume of the proposed PP-Exp and PP-MI.
We assume that the assistant server $T$ has generated enough random numbers for the PP-GPR calculation process in the offline stage
and sent their shares to the corresponding computing server a priori. 

To execute PP-Exp for an $n$-dimensional vector $\mathbf{u}$, Algorithm~\ref{alg-exp} only requires $1$ round of communication between the computing servers in Line~\ref{line:rec} of Algorithm~\ref{alg-exp}
and the amount of this communication is $2nl$. 

In PP-MI, we decompose the matrix inversion into addition, multiplication, and division operations by exploiting the Cholesky decomposition.
%
A single-element SS-based multiplication requires $1$ rounds of bidirectional communication with a traffic volume of $2l$. Implementing the division of a single element by invoking the privacy-preserving division provided in Crypten takes $17$ rounds of communication and a communication volume of $\mathcal{O}(l)$.
The PP-MI of an $n\times n$ positive definite matrix includes $n$ rounds of division and $5n - 6$ rounds of multiplication.
Thus, the total number of communication rounds for PP-MI is $17n+(5n-6) = 22n -6$.
In the PP-MI, it is necessary to perform $\mathcal{O}(n^3)$ element multiplication and $\mathcal{O}(n^2)$ element division. Therefore, the total communication volume of the PP-MI algorithm is $\mathcal{O}(n^3l)$. See Appendix C for detail.

\section{Experiments and discussion}\label{sec:experi}

This section empirically evaluates the performance of the proposed privacy-preserving operations and PP-GPR. The PP-Exp, PP-MI, and PP-GPR are built upon the open-source privacy-preserving ML framework Crypten \citep{knott2021crypten}. We set $l = 64$ and $l_f = 26$.
The experiments are carried out on three servers with a 48-core Intel Xeon CPU running at 2.9GHz and a local area network with a communication latency of 0.2ms and bandwidth of 625MBps.



\begin{figure}[t]
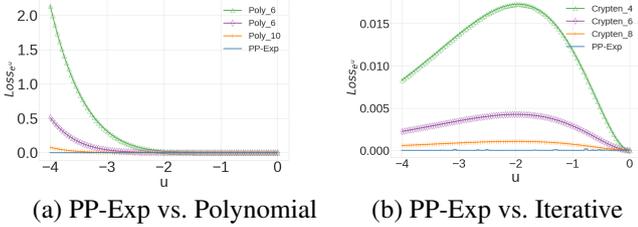

	\centering
	\begin{tabular}{cc}
		\hspace{-7mm}\includegraphics[scale=0.111]{PP-Exp_polynomial.pdf} & \hspace{-1mm}\includegraphics[scale=0.111]{PP-Exp_crypten.pdf} \\
		\hspace{-3mm}(a) PP-Exp vs. Polynomial
& (b) PP-Exp vs. Iterative\\
	\end{tabular}
	\caption{Graphs of PP-Exp vs. (a) polynomial approximation methods; and (b) the iterative approximation method.}
	\label{Fig.PP-Exp}
\end{figure}

\subsection{Evaluation of PP-Exp and PP-MI}
\label{sec:experi-ops}
We first demonstrate the accuracy and computational efficiency of the proposed operations.



\begin{table} \footnotesize
	\centering
	\caption{The computational time (in the unit of second) incurred by the tested approaches in computing $e^\mathbf{U}$ with varying size of $\mathbf{U}$.} 
	\begin{tabular}{l|cccc}
		\toprule
		\diagbox [width=6em,trim=l] {Approach}{Size of U} & $1000^2$ & $3000^2$ & $5000^2$  & $10000^2$ \\
		\hline
		Plaintext & 0.008 & 0.014 & 0.024 & 0.064   \\
		Poly\_10 & 3.308 & 20.956 & 56.462 & 221.582  \\
		Crypten\_8 & 2.557 & 11.701 & 29.528 & 113.136  \\
		PP-Exp & $\mathbf{0.208}$ & $\mathbf{0.457}$ & $\mathbf{0.935}$ & $\mathbf{3.006}$\\
		\bottomrule
	\end{tabular}
	\label{tab:PP-Exp}
\end{table}

\textbf{Evaluation of PP-Exp:} 
We compare the performance of the proposed PP-Exp against that of (a) \emph{Plaintext}: the conventional exponential operation; (b) \emph{Poly}: a polynomial approximation based on Taylor expansions; and (c) \emph{Crypten}: an iterative approach based on the limit approximation of the exponential function. 
We evaluate the accuracy and efficiency of the PP-Exp algorithm separately. For accuracy, Poly with varying degrees of polynomials and Crypten with a varying number of iterations are tested.
$Loss_{e_u}$ is the difference between the tested algorithm and $e_u$ computed in Plaintext. As shown in Fig.~\ref{Fig.PP-Exp}, the proposed PP-Exp achieves almost the same results as the plaintext and outperforms other tested algorithms.
%
Next, we compare the efficiency of PP-Exp, Poly (with polynomial degree 10), and Crypten (with 8 iterations) on varying sizes of input variables (i.e., tested on $e^{\mathbf{U}}$ with varying sizes of $\mathbf{U}$).
The computational time of the tested algorithms is shown in Table~\ref{tab:PP-Exp}.
As can be seen, the PP-Exp incurs significantly less time than both the polynomial and iterative approaches. In particular, PP-Exp is at most 70 times faster than Poly\_10 and 38 times faster than Crypten\_8 given a large size of inputs.

\begin{figure}[t]
	\centering
	\begin{tabular}{cc}		\hspace{-3mm}\includegraphics[scale=0.100]{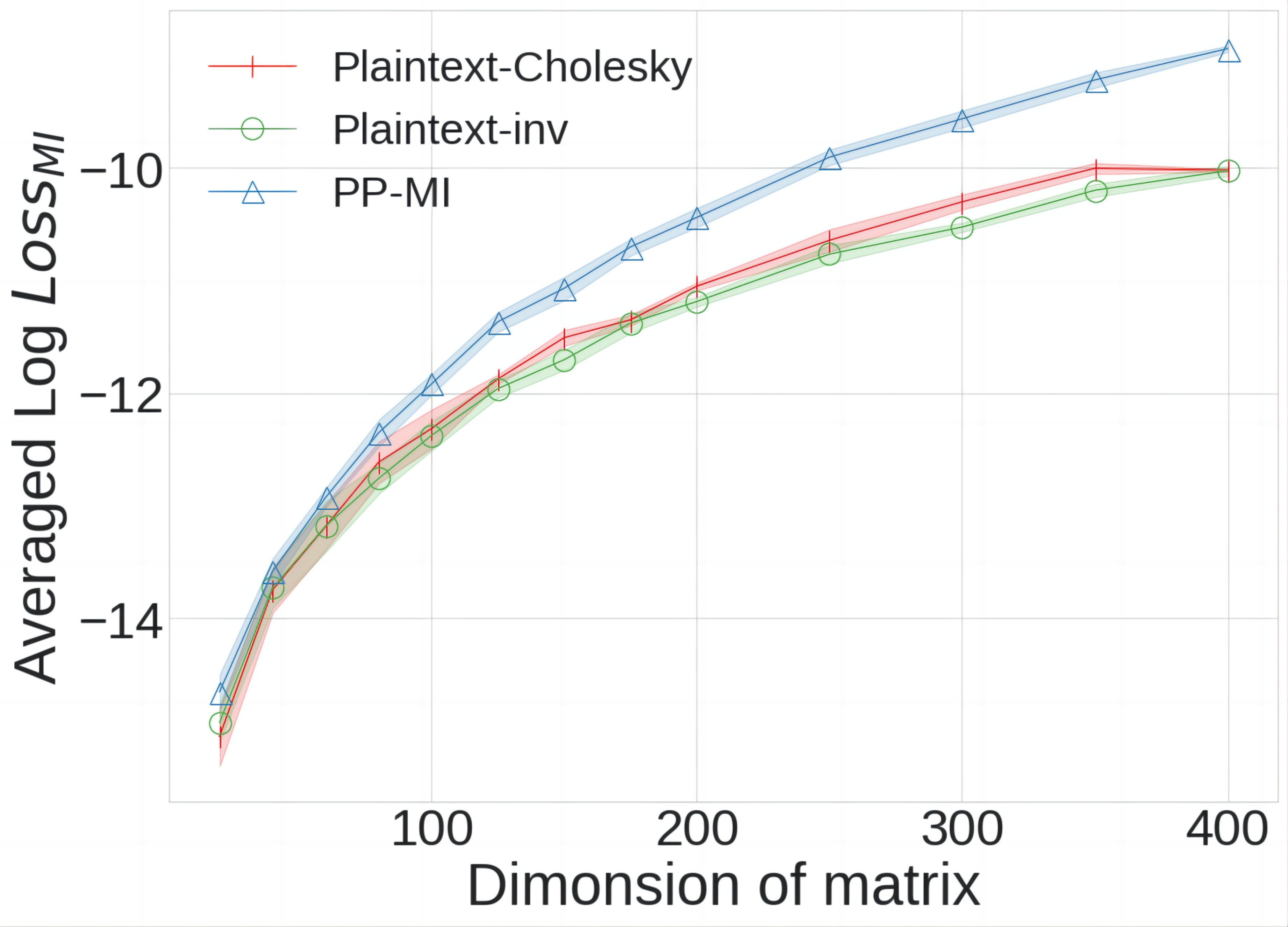} &
\hspace{-2mm}\includegraphics[scale=0.072]{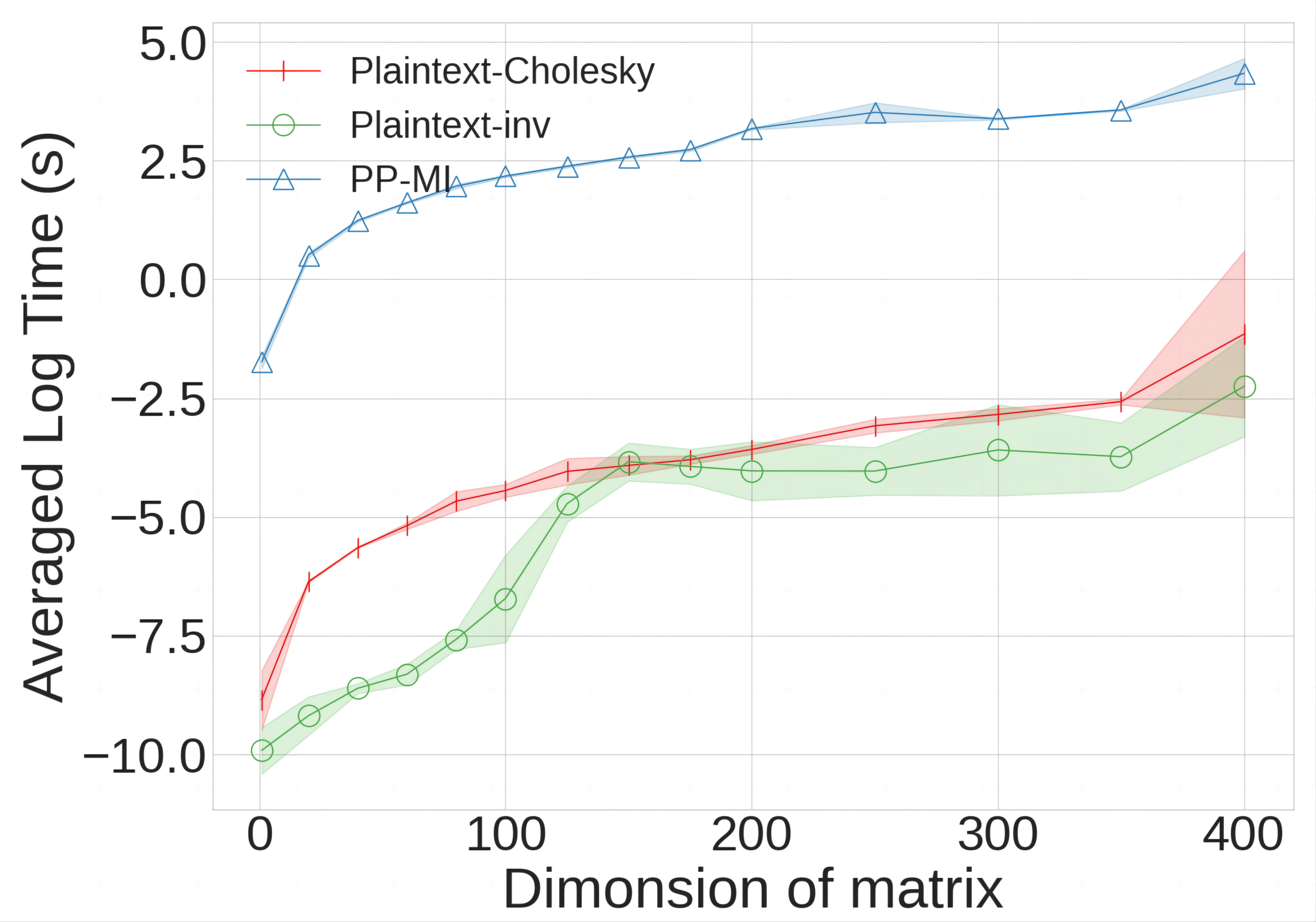}\\
\hspace{-3mm} (a) PP-MI Loss & \hspace{-3mm}(b) PP-MI Time\\
	\end{tabular} 
	\caption{Graphs of (a) Losses and (b) Computational time of different MI algorithms vs. dimension of matrix.} 
	\label{Fig.main2}
\end{figure}

\textbf{Evaluation of PP-MI:} The performance of PP-MI is tested by generating random covariance matrices and compared against that of (a) \emph{Plaintext-Cholesky}: Matrix inversion via Cholesky decomposition; and (b) \emph{Plaintext-inv}: The inv function in the torch.linalg library. 
Specifically, we first randomly sample an input matrix $\mathbf{X} \in [-10, 10]^{n\times d}$ with $d = 2$ and then compute $\mathbf{K}+\sigma^2_n\mathbf{I}$ using \eqref{kernel} with $\sigma_s^2 = 1$, $\ell = 1$, and $\sigma_n^2 = 0.1$. Let $\mathbf{\Lambda}$ be the output of a matrix inversion algorithm. $Loss_\text{MI} \triangleq ||(\mathbf{K}+\sigma^2_n\mathbf{I})\mathbf{\Lambda} - \mathbf{I}||^2_2$
is used as the inversion accuracy metric.
The $Loss_\text{MI}$ and wall-clock time of the tested algorithms averaged over 10 independent runs with varying $n$ are shown in Fig.~\ref{Fig.main2}. The error bars are computed in the form of standard deviation.
As can be seen, PP-MI incurs an acceptable level of accuracy loss (around $0.0001$ for $n = 400$) with acceptable computational cost.  
This loss comes from the approximation of the SS-based division and the fixed-point encoding steps which cannot be avoid in most SS-based algorithms.


\begin{table*} \footnotesize
\centering
\caption{Evaluation results of GPR and PP-GPR using RBF kernel with varying sizes of observations and test inputs.}
\begin{tabular}{c|c|c|c|c|c|c}
\toprule
          & \multicolumn{2}{c|}{ Dataset Size} & $Loss_\mu$ & $Loss_{\sigma^2}$ & \multicolumn{2}{c}{Time (s)}  \\ 
 \cline{2-3}  \cline{6-7}  &&&&&\\[-0.8em]
          & $n$  & Test  &  $mean (std.)$                           & $mean (std.)$ & GPR & PP-GPR      \\ 
\hline &&&&&&\\[-0.8em]
          & 80        &  20             & 0.0005\%($\pm$7.5e-05)                & 0.0141\%($\pm$3.7e-03)         & 0.028 &7.068        \\
Traffic & 150     & 50      & 0.0027\%($\pm$1.0e-02)             & 0.0061\%($\pm$4.9e-04)         & 0.089    & 13.016       \\
          & 300      & 100      & 0.0057\%($\pm$3.3e-03)         & 0.0852\%($\pm$1.2e-02)        & 0.149   & 32.355      \\ 
\hline &&&&&&\\[-0.8em]
        & 80         & 20              &  0.0007\%($\pm$3.6e-04)             & 0.0095\%($\pm$2.2e-03)       & 0.025   & 7.024       \\
Diabetes & 150        & 50               & 0.0018\%($\pm$1.3e-03)              & 0.0059\%($\pm$1.2e-03)         & 0.104    & 13.901        \\
          & 300         & 142       & 0.0058\%($\pm$2.5e-03)                 & 0.0848\%($\pm$5.6e-03)        & 0.671    & 97.076        \\
\bottomrule
\end{tabular}\label{tab:real}
\end{table*}

\begin{table*} \footnotesize
\centering
\caption{Evaluation results of GPR and PP-GPR using Mat{\'e}rn kernel with varying sizes of observations and test inputs.}
\begin{tabular}{c|c|c|c|c|c|c}
\toprule
          & \multicolumn{2}{c|}{ Dataset Size} & $Loss_\mu$ & $Loss_{\sigma^2}$ & \multicolumn{2}{c}{Time (s)}  \\ 
 \cline{2-3}  \cline{6-7}  &&&&&\\[-0.8em]
          & $n$  & Test  &  $mean (std.)$                           & $mean (std.)$ & GPR & PP-GPR      \\ 
\hline &&&&&&\\[-0.8em]
          & 80        &  20             &  
          0.2711\%($\pm$2.3e-02)             & 0.0221\%($\pm$5.5e-06)       & 0.032 & 9.732      \\
Traffic & 150     & 50      & 0.2665\%($\pm$6.1e-03)     & 0.0241\%($\pm$1.2e-06)         & 0.078       & 13.116  \\
          & 300      & 100      & 0.8288\%($\pm$1.5e-02)   & 0.0257\%($\pm$2.0e-06) &   0.153  & 35.33  \\ 
\hline &&&&&&\\[-0.8em]
        & 80         & 20              &   0.0548\%($\pm$ 1.0e-03)& 0.0193\%($\pm$4.0e-06) & 0.023  &  8.027        \\
Diabetes & 150        & 50  &              0.0424\%($\pm$6.2e-05) & 0.0236\%($\pm$ 7.6e-06)& 0.102   & 15.702      \\
          & 300         & 142       & 0.0545\%($\pm$4.9e-06)& 0.0343\%($\pm$ 3.0e-06)& 0.589   & 99.082 \\
\bottomrule
\end{tabular}\label{tab:matern}
\end{table*}

\subsection{Evaluation of PP-GPR}
This section empirically evaluates the performance of the proposed PP-GPR on two 
real-world datasets: (a) \emph{Traffic} dataset \citep{chen2015gaussian} contains taxi demand information of 2506 regions in a city between 9:30 p.m. and 10 p.m. on August 2, 2010; and (b) \emph{Diabetes} dataset (under BSD License) \citep{efron2004least} contains diabetes progression of 442 diabetes patients with 10 input features.
We test the proposed PP-GPR with both the SE kernel \eqref{kernel} and the Mat{\'e}rn$_{3/2}$ kernel:
\begin{equation*}
k(\mathbf{x}, \mathbf{x}') \triangleq \sigma^2_s(1 + \sqrt{3d(\mathbf{x}, \mathbf{x}')}/l)\text{exp}(-\sqrt{3d(\mathbf{x}, \mathbf{x}'})/l)
\end{equation*}
In the diabetes experiments, we use $\sigma^2_s = 0.8 $, $\sigma^2_n = 0.1 $, and $\ell = 0.23$ for the SE kernel and $\sigma^2_s = 0.1 $, $\sigma^2_n = 0.1 $, and $\ell = 1.0$ for the Mat{\'e}rn$_{3/2}$ kernel. As the traffic dataset suggested, we set $\sigma^2_s = 0.1 $, $\sigma^2_n = 0.1 $, and $\ell = 1.0$ for all the traffic experiments. 
%

Let $\mathcal{X_*}$ be a set of test inputs, $\mu_{\mathbf{x}_*|\mathcal{D}}$ ($\sigma^2_{\mathbf{x}_*|\mathcal{D}}$) and $\Tilde{\mu}_{\mathbf{x}_*|\mathcal{D}}$ ($\Tilde{\sigma}^2_{\mathbf{x}_*|\mathcal{D}}$) be, respectively, the predictive mean (variance) of the GPR and PP-GPR. The relative difference between the predictive results of GPR and PP-GPR is used as the performance metric: $Loss_\mu \triangleq |\mathcal{X}_*|^{-1}\sum_{\mathbf{x}_* \in \mathcal{X}_*}(|\mu_{\mathbf{x}_*|\mathcal{D}} - \Tilde{\mu}_{\mathbf{x}_*|\mathcal{D}}|/\mu_{\mathbf{x}_*|\mathcal{D}})$. $Loss_{\sigma^2}$ is computed in a similar way. 
To test the performance of PP-GPR in different data scales, we randomly sample observations and test data from each dataset with varying $n$ and $|\mathcal{X}_*|$.
%
The loss of the predictive results and the wall-clock execution time (including both computation and communication time) are shown in Table~\ref{tab:real} and Table~\ref{tab:matern}. All the results are averaged over 5 random runs.

It can be observed that the PP-GPR achieves a similar predictive mean and variance compared to conventional GPR. The losses are due to the approximation of some SS-based operations (e.g., division) and the fixed-point encoding step.
The computational errors of the Mat{\'e}rn kernel are slightly higher than that of the SE kernel but still remain at a low level. Further analysis revealed that the higher computational error in the Mat{\'e}rn kernel is due to the inclusion of the expression $\sqrt{d(\mathbf{x}, \mathbf{x}^\prime)}$. The square root operation is a non-linear operation that must be approximated using the Newton iterative approach in Crypten~\citet{knott2021crypten}, which results in the higher computational error. In our future work, we plan to conduct further research to investigate this issue.

Furthermore, although PP-GPR incurs a longer time than GPR, especially if $n$ is large, it can finish the model construction and prediction in a reasonable time ($< 2$ mins) for a dataset with several hundred observations.


In Appendix D, we perform an additional empirical comparison between our algorithm and DP-based GPR under the scenario that only the model outputs are sensitive.
We believe that this comparison is fair, given that both methods can theoretically preserve privacy.
The other privacy-preserving GPR approaches (e.g., FHE-based and FL-based GPR) are not compared since even when operating within the same scenario (i.e., HDS, VDS, or PDS), they may have fundamentally different security assumptions to that of PP-GPR, which ultimately makes them incomparable. See Section~\ref{sec:related} for detailed discussions.
 

\section{Related work}\label{sec:related}


To the best of our knowledge, there is no existing PP-GPR work that is designed based on SMPC techniques. As has been mentioned in Section~\ref{sec:intro}, although some other privacy enhancement techniques have been applied to GPR, none of them is practical enough to protect the privacy of both the inputs and outputs of GPR for all the three data-sharing scenarios (i.e., HDS, VDS, and PDS).
To be specific, \citet{fenner2020privacy} considers only the PDS scenario and protects the input features of the test data by \emph{fully homomorphic encryption} (FHE) algorithm. Since performing computation on the homomorphically encrypted data incurs high computational costs, they do the PP-GPR prediction through interactive calculations between the user and the model constructor. Such an interactive method, however, cannot be generalized to FHE-based PP-GPR model construction step since the covariance matrix inversion operation is not considered.

Another technique that is widely used to achieve PP-ML models is \emph{differential-privacy} (DP). \citet{smith2018differentially} proposed the first DP-GPR algorithm which can only protect the privacy of the model outputs~$\mathbf{y}$. \cite{kharkovskii2020private} proposed a DP method to protect the input features of the GPR model via random projection. However, this method requires all the observations used for GPR model construction to belong to a single curator and thus, cannot be applied to either HDS or VDS scenarios. In addition, the DP-based method may incur large DP noise to the original model when the privacy budget $\epsilon$ is small, which may significantly reduce the model performance \citep{dwork2014algorithmic}.

Some other works~\citep{dai2020federated,kontoudis2022fully,yue2021federated} consider protecting the privacy of the GPR observations via \emph{federated learning} (FL) or combine FL with DP to further protect the privacy of the model parameters~\citep{dai2021differentially}. To convert the GPR model construction into a distributed/federated manner, these works have to exploit some sparse approximations (e.g., random features) of the conventional GPR, which may reduce the model performance. Moreover, FL-based GPR works can only be applied to the HDS scenario.   

Recently,~\cite{kelkar2021secure} developed a privacy-preserving exponentiation algorithm based on secret sharing techniques in a two-server setting. The communication overhead of this algorithm in the online phase is comparable to that of PP-Exp (i.e., one round of communication and transmission of two elements). However, the algorithm requires an expensive cryptographic primitive (i.e., homomorphic encryption) in the preprocessing phase to generate the random numbers needed in the online phase, resulting in excessive overhead. In addition, the algorithm suffers a certain probability of error from the use of a secure ring change procedure. In the setting of this paper (i.e., $l=64, l_f =26$), the probability that the error occurs is $\frac{1}{4}$, which is unacceptable.

Note that even when operating within the same scenario (i.e., HDS, VDS, or PDS), different privacy-preserving approaches may have fundamentally different security assumptions~\citep{yin2021comprehensive, zhang2022no}. 
Specifically, the privacy of the FHE-based GPR algorithm~\citet{fenner2020privacy} may be at risk even in the PDS scenario due to the decryption steps designed for reducing the high computational cost of the exponential operation in FHE. However, this work provides a solution by effectively addressing the challenges posed by the SS-based exponentiation operation. Consequently, this work guarantees complete privacy protection across the entire PDS process.
The FL-based GPR~\citep{dai2020federated,kontoudis2022fully,yue2021federated} has no theoretical analysis of its privacy-preserving capabilities. This is because the intermediate results (e.g., local model parameters or gradients) generated during the algorithm need to be exchanged between the server and clients. Numerous studies~\citep{zhu2019deep, zhao2020idlg} have demonstrated that these intermediate results pose a potential risk of revealing private data.

\section{Conclusion}\label{sec:conclusion}

This paper describes the first SS-based privacy-preserving GPR model which can be applied to both horizontal and vertical data-sharing scenarios. We provide a detailed workflow for implementing both the model construction and prediction steps of PP-GPR. Two additive SS-based operations (i.e., PP-Exp and PP-MI) are proposed such that they can be combined with existing SS-based operations for constructing a secure and efficient GPR model. We analyze the security and computational complexity of the proposed operations in theory.
Although PP-GPR incurs more computational time due to additional communications between the two computing servers and some additional computing steps, it can perform GPR in an acceptable time with a security guarantee, which is a superior alternative to existing FL and DP-based privacy-preserving GPR approaches when the scale of observations is not large.
%

\section{Acknowledgements}\label{sec:Acknowledgements}
This research is partially supported by the National Key Research and Development Program of China (No. 2022ZD0115301), the National Natural Science
Foundation of China (No. 62206139), and a key program of fundamental research from Shenzhen Science and Technology Innovation Commission (No. JCYJ20200109113403826).

\bibliography{luo_732}

\onecolumn 

\appendix

\section{Correctness and Security analysis of PP-Exp}\label{app:ppexp}

\subsection{Proof of Theorem 1}
In terms of algorithm correctness, we need to consider the problem of overflow or underflow errors during the execution of Algorithm 2. First, the PP-Exp algorithm involves computations on two algebraic structures, namely, the ring of integers $\mathcal{Z}_L$ modulo $L$ and the fixed-point set $\mathcal{Q}_{<\mathcal{Z}_L,l_f>}$. Specifically, the operations on shares are performed on $\mathcal{Z}_{L}$ and others are performed on $\mathcal{Q}_{<\mathcal{Z}_L,l_f>}$. 

\noindent{\textbf{Underflow:}} The PP-Exp has the risk of underflow when calculating $e^{-\check{r}}$ and $e^{\check{d}}$ in Line 5 and Line 13 of Algorithm~2. The minimum value of $e^{-\check{r}}$ and $e^{\check{d}}$ is $e^{u_{min} -\check{r}_{max}}$. To guarantee that PP-Exp will not overflow during computation, we need ensure that $e^{u_{min} -\check{r}_{max}}$ can be expressed as a fixed-point number with precision $l_f$. This is equivalent to $e^{u_{min} -\check{r}_{max}} \cdot 2^{l_f}\geq1$. Therefore, $l_f\geq (\check{r}_{max} - u_{min})\log_2^e$ is needed to ensure that PP-Exp does not underflow during the calculation.

\noindent{\textbf{Overflow:}}
PP-Exp has the potential to overflow when computing $[e^u]_j$ for $j \in \{0,1\}$ in Line 14. Since $[e^u]_j$ is the share of $e^u$, the maximum value of the reconstruction process is
$$
[e^{u_{max}}]_0 +  [e^{u_{max}}]_1 = e^{u_{max} + \check{r}}\cdot 2^{l_f} \cdot ([e^{-\check{r}}]_0 + [e^{-\check{r}}]_1) = e^{u_{max}} \cdot 2^{2l_f}
$$
according to Line 14 of Algorithm 2. Therefore, $ e^{u_{max}} \cdot 2^{2l_f} \leq 2^{l-1}$ is required to ensure that the reconstruction process does not overflow. As $u<0$ (i.e., $u_{max} = 0$), it is achieved that PP-Exp does not overflow if $l_f< \frac{l-1}{2}$.

\subsection{Proof of Theorem 2}
In terms of algorithm security, we analyze the probability of PP-Exp privacy leakage and the \textit{expected degree of information leakage}.
The number of possible values of $u$ is $m_u$ and the number of possible values of $r$ is $m_r$, then the probability that the algorithm leads to additional privacy leakage of input $u$ is
\begin{equation}
	\frac{(1+2+\cdots+m_u)\cdot2}{m_um_r}=\frac{m_u\cdot (m_u-1)}{m_um_r}=\frac{m_u-1}{m_r}.
\end{equation}
The probability that the PP-Exp is \emph{secure} is $1 - \frac{m_u-1}{m_r} = \frac{m_r-m_u+1}{m_r}$.

Since the exponents are all negative in the Gaussian process regression, the number of $u$ values is reduced by half. This allows the probability of input $u$ leakage to be further reduced. For example, when $l_f = 2^{29}$, supposing the input $u$ takes values in the range $[-4,0]$ and $r$ takes values in the range $[-16,16)$ (i.e., the number of values of $u$ and $r$ are $2^{31}+1$ and $2^{34}+1$, respectively), the probability that the algorithm to be secure is 
$\frac{2^{34}-2^{31}}{2^{34}} =\frac{7}{8}$.

The \emph{expected degree of information leakage} is used to describe the amount of privacy leakage of the PP-Exp algorithm. It can be understood by the following simple example. Suppose $u \in \{-2,-1,0\}, r \in \{-2,-1,0, 1, 2\}$ and $d = u+r \in \{-4, -3, -2,-1, 0, 1, 2\}$. The probabilities of $d$ taking different values are as follows: 
\begin{equation}
	\begin{array}{cc}
		&  \Pr\{d = -4\} = \Pr\{d = 2\} = \frac{1}{15},\\
		& \Pr\{d = -3\} = \Pr\{d = 1\} = \frac{2}{15},\\
		&  \Pr\{d = -2\} = \Pr\{d = -1\} = \Pr\{d = 0\}=\frac{3}{15}.   
	\end{array}  
\end{equation}
If $d\in \{-2, -1, 0\}$, $u$ can take any of $-2, -1, 0$ such that no additional information about $u$ is revealed by $d$. Therefore, the degree of information leakage of $u$ is $\frac{3}{15}\cdot \frac{1}{3} = \frac{1}{5}$. If $d\in \{-4, 2\}$, $u$ will only correspond to unique values, and the degree of information leakage of $u$ is $2\cdot \frac{1}{15}\cdot 1$.  If $d\in \{-3, 1\}$, there are two possible values of $u$. The degree of information leakage of $u$ is $2 \cdot\frac{2}{15} \cdot \frac{1}{2} = \frac{2}{15}$. Therefore,  the \textit{degree of information leakage} of $u$ averaged over $d \in \{-4, -3, 1, 2\}$ is $\frac{2}{15} + \frac{2}{15} + \frac{1}{5} = \frac{7}{15}$. 

In PP-Exp, there are $\frac{2k}{m_um_r}, k = 1,2,\cdots, m_u-1$ probabilities that $m_u-k$ possible values of the input $u$ are excluded by $d$. Therefore, the \textit{average degree of information leakage} of input $u$ is 
\begin{equation}
	(1-\frac{m_u -1}{m_r})\cdot\frac{1}{m_u} + \sum_{k = 1}^{m_u-1} (\frac{2k}{m_um_r})\cdot\frac{1}{k} = \frac{m_r -m_u +1}{m_um_r} + \frac{2(m_u-1)}{m_um_r} = \frac{m_u+m_r-1}{m_u\cdot m_r}.
\end{equation}
It is important to note that the PP-Exp algorithm only exposes the exact value of $u$ when the maximum and minimum values are in the range where $u$ and $r$ are taken simultaneously. This probability is 
\begin{equation}
	\frac{2}{mn}=\frac{2}{(2^{31}+1)\cdot (2^{34}+1)}<\frac{2}{2^{31}\cdot 2^{34}}=\frac{1}{2^{64}}.
\end{equation} 


\begin{algorithm}[b]\footnotesize
	\caption{Privacy-preserving Cholesky decomposition}
	\label{alg2}
	\textbf{Setup.} The servers determine an integer ring $\mathcal{Z}_L$.\\
	\textbf{Input.} 
	$S_0$ holds the share $[\mathbf{U}]_0$; $S_1$ hold the share $[\mathbf{U}]_1$.
	\begin{algorithmic}[1]
		\State 	// \textbf{Offline phase:}
		\State $T$ generates $\mathbf{A}, \mathbf{B} \in Z^{n\times n}_L$ randomly, and calculates $\mathbf{C} = \mathbf{AB}$;
		\State $T$ sends $([\mathbf{A}]_j, [\mathbf{B}]_j,[\mathbf{C}]_j)$ to $S_j$.
		\State // \textbf{Online phase:}
		\For{$j \in \{0,1\}$}:
		\State $[d_1]_j \leftarrow [u_{1,1}]_j$;
		
		\State $S_j$ calculates $[l_{1:n,1}]_j = [u_{1:n, 1}]_j / [u_{1,1}]_j$ by calling PP-Div;
		\For{$k \in \{2,3,\ldots,n\}$}:
		\State $S_j$ calculates $[d_k]_j = [u_{k,k}]_j - [\sum_{m = 1}^{k-1}l^{2}_{k,m}d_m]_j$ by calling  PP-MM;
		\State $S_j$ calculates $[\hat{l}_{k+1:n,k}]_j =  [u_{k+1:n,k}]_j - [\sum_{m = 1}^{k-1}l_{k,m}d_ml_{k+1:n,m}]_j$ by calling  PP-MM;
		\State $S_j$ calculates $[l_{k+1:n,k}]_j = [\hat{l}_{k+1:n,k}]_j / [d_k]_j$ by calling  PP-Div;
		\EndFor
		\EndFor
		\Comment $S_j$ gets $[\mathbf{L}]_j, [\mathbf{D}]_j$.
	\end{algorithmic}
	\textbf{Output.} 
	$S_j$ output the share $[L]_j, [D]_j$ for $j \in \{0,1\}$.
\end{algorithm}

\section{Privacy-preserving matrix inversion}\label{app:ppmi}
In this section, we will give the construction details of the \emph{privacy-preserving matrix inversion} (PP-MI) operation.
The main idea of the PP-MI is to transform the matrix inversion process into MPC-friendly operations such as multiplication and division.
Next, we will first briefly recall the process of positive definite matrix inversion via Cholesky decomposition and then provide the 
pseudo-code of the PP-MI algorithm. 


Let $\mathbf{U} = (u_{h, k})_{h, k = 1, 2, \ldots, n}$ be an $n\times n$ positive definite matrix. It can be decomposed as $\mathbf{U} = \mathbf{LDL}^\top$ where $\mathbf{L}$ is a unit lower triangular matrix and $\mathbf{D}$ is a diagonal matrix.
Let $d_k$ be the $k^{th}$ diagonal element of $\mathbf{D}$ for $k = 1, \ldots, n$. 
We can calculate $\mathbf{L},\mathbf{D}$ through \eqref{eq4}. 
For $k = 1,2,\dots, n$ and $h = k+1, k+2, \dots, n$.
\begin{equation}
	\label{eq4}
	\left\{
	\begin{aligned}
		& d_k = u_{k,k} - \sum_{m = 1}^{k-1}l^{2}_{k,m}\cdot d_m\ , \\
		& l_{h,k} = (u_{h,k} - \sum_{m = 1}^{k-1}l_{h,m}\cdot l_{k,m}\cdot d_m)/ d_k\ .
	\end{aligned}
	\right.
\end{equation} 

Supposing $\mathbf{U\Lambda} = \mathbf{I}$ for a given $n \times n$ identity matrix $\mathbf{I}$. Next, we will introduce how to compute $\mathbf{\Lambda}$ from $\mathbf{I, L}$, and $\mathbf{D}$. Then, we can have $\mathbf{U}^{-1} = \mathbf{\Lambda}$.  

Let $\mathbf{V} \triangleq \mathbf{DL}^\top\mathbf{\Lambda}$. Then, $\mathbf{U\Lambda} = (\mathbf{LDL}^\top)\mathbf{\Lambda} = \mathbf{LV} = \mathbf{I}$ and $\mathbf{V}$ is a unit lower triangular matrix. For $k = 1, 2,\dots, n$ and $h = k+1, k+2,\dots, n$, we can calculate each element of $\mathbf{V}$ as:
\begin{equation}
	\label{eq5}
		v_{h,k} = -\sum_{m=1}^{h-1}v_{m,k}\cdot l_{h,m}.
	\end{equation} 
	
	we can calculate the matrices $\mathbf{\Lambda}$ using \eqref{eq6}:
	\begin{equation}
		\label{eq6}
		\mathbf{\Lambda} = \mathbf{U}^{-1} = (\mathbf{(LDL^\top)})^{-1} = (\mathbf{L}^{-1})^\top \mathbf{D}^{-1} \mathbf{L}^{-1} = \mathbf{V}^\top \mathbf{D}^{-1} \mathbf{V}.
	\end{equation}

	
	Based on equations \eqref{eq4}, \eqref{eq5}, and \eqref{eq6}, we can implement PP-MI by reasonably invoking privacy-preserving multiplication (PP-MM) and privacy-preserving division (PP-Div) in \cite{knott2021crypten}. Let $\mathbf{U}$ be the an $n \times n$ positive definite matrix, $u_{a;b, k:h} \triangleq (u_{i, j})_{i=a, \ldots, b}^{j = k, \ldots, h}$. The PP-MI algorithm for $\mathbf{U}$ consists of three parts, and their pseudo-codes refer to the Algorithms~\ref{alg2}-\ref{alg4}. Note that the pseudo-codes are note exactly the same as \eqref{eq4}-\eqref{eq6} since we vectorize some steps for accelerating the computation. 
	
	
	\begin{algorithm}[t]\footnotesize
		\caption{Privacy-preserving forward}
		\label{alg3}
		\textbf{Setup.} The servers determine an integer ring $\mathcal{Z}_L$.\\
		\textbf{Input.}
		$S_0$ holds the share $[\mathbf{U}]_0, [\mathbf{L}]_0$; $S_1$ hold the share $[\mathbf{U}]_1,[\mathbf{L}]_1$.
		\begin{algorithmic}[1]
			\State 	// \textbf{Offline phase:}
			\State $T$ generates $\mathbf{A}, \mathbf{B} \in Z^{n\times n}_L$ randomly, and calculates $\mathbf{C} = \mathbf{AB}$;
			\State $T$ sends $([\mathbf{A}]_j, [\mathbf{B}]_j,[\mathbf{C}]_j)$ to $S_j$.
			\State // \textbf{Online phase:}
			\State $v_{1, 1} \leftarrow 1$
			\For{$j \in \{0,1\}$}
			\For{$k \in \{2,3,\dots,n\}$}
			\State $S_j$ calculates $[v_{k,1:k-1}]_j =  - [\sum_{m = 1}^{k-1}v_{m,1:k-1}l_{k,m}]_j$ by calling  PP-MM;
			\EndFor
			\EndFor
			\Comment $S_j$ gets $[\mathbf{V}]_j$.
		\end{algorithmic}
		\textbf{Output.}
		$S_j$ outputs the share $[\mathbf{V}]_j$ for $j\in \{0,1\}$.
	\end{algorithm}

	\begin{algorithm}[t]\footnotesize
		\caption{Privacy-preserving backward}
		\label{alg4}
		\textbf{Setup.} The servers determine an integer ring $\mathcal{Z}_L$.\\
		\textbf{Input.}
		$S_0$ holds the share $[\mathbf{V}]_0, [\mathbf{L}]_0, [\mathbf{D}]_0$; $S_1$ hold the share $[\mathbf{V}]_1,[\mathbf{L}]_1, [\mathbf{D}]_1$.
		\begin{algorithmic}[1]
			\State // \textbf{Offline phase:}
			\State $T$ generates $\mathbf{A}, \mathbf{B} \in Z^{n\times n}_L$ randomly, and calculates $\mathbf{C} = \mathbf{AB}$;
			\State $T$ sends $([\mathbf{A}]_j, [\mathbf{B}]_j,[\mathbf{C}]_j)$ to $S_j$.
			\State // \textbf{Online phase:}
			\State $S_j$ calculates $[\mathbf{\Lambda}]_j = [\mathbf{V}^\top]_j [\mathbf{D}^{-1}]_j [\mathbf{V}]_j$ by calling PP-Div and PP-MM.
			\Comment $S_j$ gets $[\mathbf{\Lambda}]_j$.
		\end{algorithmic}
		\textbf{Output.}
		$S_j$ outputs the share $[\mathbf{\Lambda}]_j$ for $j \in \{0,1\}$.
	\end{algorithm}

	
	\section{Communication complexity analysis}\label{app:proof}
	
	In this section, we will analyze the communication complexity of PP-MI step by step. We only analyze the communication complexity in the online phase. The process of the offline phase can be done by the server during idle time. 
	
	
	
	For PP-MI, we will analyze the communication cost for Algorithms~\ref{alg2},~\ref{alg3}~and~\ref{alg4}, separately. For Algorithm~\ref{alg2}, there is $1$ vector-wise division in line 7. For each $k = 2, ..., n-1$, there are $4$ rounds of vector-wise multiplication for lines 9-10 and $1$ vector-wise division for line 11. When $k = n$, there are only $2$ rounds of vector-wise multiplication for line 9. We do not need to compute lines 10-11 when $k = n$. There are $17$ rounds of communication for each element/vector-wise PP-Div and $1$ round of communication for each element/vector-wise PP-MM. Consequently, the total communication round for Algorithm~\ref{alg2} is $17 + (n-2)*(4+17) + 2 = 21n-23$.
	%
	Since there are $\mathcal{O}(n^3)$ element-wise multiplications and $\mathcal{O}(n^2)$ element-wise divisions included in Algorithm~\ref{alg2}, its communication complexity is $\mathcal{O}(n^{3}l)$.
	
	For Algorithm~\ref{alg3}, it includes $1$ vector-wise multiplication for each $k = 2, \ldots, n$
	and each round of computation (i.e., line 8 for each $k$) consists of $(k-1)^2$ element-wise multiplications. Therefore,
	Algorithm~\ref{alg3} incurs a total of $n-1$ rounds of communication and its communication complexity is $\mathcal{O}(n^{3}l)$.
	
	For Algorithm~\ref{alg4}, it contains $1$ matrix division and 1 matrix multiplication with $n^2$ elements in line 5. Therefore, the total number of communication rounds required to execute Algorithm~\ref{alg4} is $17+1=18$. The communication complexity is $\mathcal{O}(n^{2}l)$.
	
	In summary, the PP-MI algorithm incurs $21n-23+n-1+18=22n-6$ rounds of communication and the communication complexity is $\mathcal{O}(n^{3}l)$.

	\section{Additional Experimental Results (PP-GPR vs. DP-GPR)}
	
	Another technique that has been widely used to achieve privacy preservation in machine learning is \emph{differential-privacy}~(DP)~\citep{dwork2014algorithmic}.
	By adding noise that satisfies the privacy budget, DP-based privacy-preserving machine learning algorithms achieve some privacy protection at the expense of the accuracy of the model. In the domain of Gaussian process regression (GPR) algorithms,~\citet{smith2018differentially} proposed a significant contribution by presenting the first privacy-preserving GPR algorithm. However, it is important to note that their approach only safeguards the privacy of the model outputs $\mathbf{y}$ through the addition of noise levels that adhere to a given privacy budget. Consequently, their algorithm falls short of achieving complete protection for both the model inputs and outputs.
	
	To demonstrate the effectiveness of our algorithm, we present the loss of predictive mean of DP-GPR on the diabetes dataset for different levels of DP guarantee (i.e., varying $\epsilon$) using the SE kernel. As can be seen, even with a large privacy budget (i.e., $\epsilon = 1.0$), DP-GPR incurs significantly larger computational errors due to the additional DP noise compared to the proposed PP-GPR. The $Loss_{\mu}$ of our proposed algorithm is kept below $0.01\%$ (Table 2 in the main paper). 
	
	\begin{table*} [h]\footnotesize
		\centering
		\caption{Comparison of our algorithm with the DP-based algorithm in the relative difference $Loss_{\mu}$.}
		\begin{tabular}{c|c|c|c|c|c}
			\toprule
			$n$  & Test & $\epsilon = 1.0$  & $\epsilon = 0.5$  & $\epsilon = 0.2$ & \textbf{Ours}  \\ 
			
			\hline 
			80        &  20  &           $18.6\%(\pm23.4)$ & $42.9\%(\pm188.3)$ & $96.6\%(\pm584.8)$  & $0.0007\%(\pm3.6e-04)$ \\
			150     & 50   & $27.8\%(\pm19.4)$ & $61.2\%(\pm311.6)$ & $148.6\%(\pm1457.7)$ & $0.0018\%(\pm1.3e-03)$ \\
			300      & 100      &$18.3\%(\pm10.5)$ & $33.5\%(\pm23.7)$  & $80.6\%(\pm133.3)$&$0.0058\%(\pm2.5e-03)$  \\
			\bottomrule
		\end{tabular}\label{tab:dp-gp}
	\end{table*}
	
	
\end{document}